\documentstyle[12pt]{article}

\begin{document}

\date{}
\title{On the Action Principle in Affine Flat Spaces with Torsion}

\author{
{\normalsize P.~P.~Fiziev} \thanks{
Work partially supported by
the Sofia University Foundation
for Scientific Researches, Contract~No.3052/95,
and by
the National Foundation for Scientific Researches,
Contract~No.F318/95}\\
{\footnotesize Department of Theoretical Physics, Faculty of Physics,
Sofia University,}\\
{\footnotesize Boulevard~5 James Bourchier, Sofia~1126, Bulgaria }\\
{\footnotesize  E-mail: fiziev@phys.uni-sofia.bg}
}

\maketitle

\begin{abstract}
To comply with the equivalence principle in Einstein-Cartan-like
theories of gravity we propose a modification of the action principle
in affine flat spaces with torsion.
\end{abstract}


\sloppy
\renewcommand{\baselinestretch}{1.3} %
\newcommand{\sla}[1]{{\hspace{1pt}/\!\!\!\hspace{-.5pt}#1\,\,\,}\!\!}
\newcommand{\db}{\,\,{\bar {}\!\!d}\!\,\hspace{0.5pt}}
\newcommand{\partb}{\,\,{\bar {}\!\!\!\partial}\!\,\hspace{0.5pt}}
\newcommand{\dsla}{\partb}
\newcommand{\eql}{e _{q \leftarrow x}}
\newcommand{\eqr}{e _{q \rightarrow x}}
\newcommand{\ite}{\int^{t}_{t_1}}
\newcommand{\itz}{\int^{t_2}_{t_1}}
\newcommand{\itd}{\int^{t_2}_{t}}
\newcommand{\lfrac}[2]{{#1}/{#2}}

\section{Introduction}
The affine-metric geometry has got many physical applications in
different domains. Here is a list of part of them:

1.The theory of the Kustaanheimo-Stiefel transformation in
celestial mechanics \cite{KuSt},\cite{StSch},
and the corresponding extension in
quantum mechanics, especially, the Duru-Kleinert transformation
in calculation of the Feynman path integral for Colomb potential
(see for example \cite{Kleinert1} and the references there in).

2.The theory of the plastic deformations in solid states
(see for example \cite{Schouten} -- \cite{Kleinert2},
and the references there in).

3.In most of the modern generalizations of the general relativity
(GR) like:

i)Einstein-Cartan (EC) theories (see for example \cite{Cartan}
-- \cite{SabGas}, and the references there in);

ii)gauge theories of gravity -- Einstein-Cartan-Sciama-Kible
(ECSK) theories (see for example \cite{Hehl2} -- \cite{IvaProSar},
and the references there in);

iii)affine-metric theories of gravity (see for example the review
article \cite{HeCrMiNe}, and the huge amount of references there in).

4.In the theory of "strong gravity" (see for example \cite{SabGas},
and the references there in).

5.In the theory of supergravity (see for example \cite{SabGas},
\cite{West}, and the references there in).

6.In all kinds of modern string theories .

8.In the theory of space-time defects (see for example
\cite{GaLet} -- \cite{Anandan}, and the references there in).

9.In the theory of the gravitational singularities \cite{Esposito1},
\cite{Esposito2},
\\
and so on. There exist a huge amount of papers on these subjects
and one may find the corresponding references
in the literature, cited above.

In all theories of this type the dynamical equations
are derived using the standard form of action principle in the
configuration space, which carries an affine-metric geometry.
We believe in the action principle because:
1) It's origin is in quantum mechanics \cite{Dirac}, \cite {Feynman};
2)It was verified successfully in all domains of classical physics.

Nevertheless, an invalidity of the {\em standard application}
of this fundamental principle in spaces with torsion was discovered in
\cite{FizKl1}, in connection with Kustaanheimo-Stiefel transformation
and it's application for calculations of Feynman path integral.
There one must use some modified form of the action
principle in the corresponding space with torsion to derive the right
classical equations of motion. This form is analogous (but not identical)
to the Poincar\'e modification of the variational principle of classical
mechanics in anholonomic coordinates \cite{Poincare}, \cite{Arnold},
which can be illustrated on the well known example of rigid body
dynamics in body reference system \cite{Arnold},\cite{FizKl2}.
The relation between the Poincar\'e action principle in anholonomic
coordinates and the corresponding modification of the action principle
for classical nonrelativistic particle in an affine flat spaces with torsion
has been discussed in \cite{FizKl3}.

In present article we extend at first the new form
of action principle to field theory, especially to the theory of
scalar field in flat affine-metric space with torsion.

We find a motivation for our modification
of the classical action principle
in EC and ECSK theories of gravity
(topics 2.i;ii; in the list given above).
It seems to us that this modification will be suitable
for all kinds of modern theories of gravity
and related topics 4-9,
where a torsion presents.

\section{The Violation of the Equivalence Principle in ECSK Theories}
The affine-metric geometry on the manifold $M^{(1,3)}$ with holonomic
coordinates $(q)= \{q_{\alpha=0,1,2,3}\}$
is defined:
1) by a metrics tensor
$g_{\alpha\beta}(q)= \eta_{ab} {e^a}_\alpha(q) {e^b}_\beta(q)$,
 $g=\det||g_{\alpha\beta}||\neq 0$,
where $||\eta_{ab}||= diag\{ 1,-1,-1,-1\}$ is the flat Minkowski
metric, ${e^a}_\alpha(q)$ being the components of the tetrads,
$e=\det||{e^a}_\alpha(q)||\neq 0$, $g=-e^2$;
and
2) by an affine connection coefficients
${ \Gamma_{\alpha\beta} }^\gamma (q)$ which obey the
metric-compatibility condition
$\nabla_\alpha g_{\beta \gamma}~=~0~.$ This condition leads to the following
most general form of the affine-metric connection coefficients:
\begin{equation}
{ \Gamma_{\alpha\beta} }^\gamma =
{ \gamma \brace \alpha \beta }
- {K_{\alpha\beta} }^\gamma,
\label{eq:Gamma}
\end{equation}
where ${ \gamma \brace \alpha \beta }=
{\frac 1 2}g^{\gamma \sigma}
(\partial_\alpha g_{\beta \sigma}
+\partial_\beta g_{\alpha \sigma}
-\partial_\sigma g_{\alpha \beta})$
are the Christofel symbols,
$g^{\alpha \beta}$
being the elements of the inverse matrix to the matrix
with elements $g_{\alpha \beta}$;
${S_{\alpha\beta} }^\gamma~=~{ \Gamma_{[\alpha \beta]}}^\gamma $
are the components of the torsion tensor in holonomic coordinates, and
${K_{\alpha\beta} }^\gamma =
-{S_{\alpha\beta} }^\gamma
-{S^\gamma}_{\alpha\beta}
-{S^\gamma}_{\beta\alpha} $
are the components of the contorsion tensor.

The EC theories are based on the idea, which seems to be very deep.
In a flat Minkowski space $E^{(1,3)}$ the physics must be invariant with
respect to the Poincar\'e group ${\cal P}(1,3)$ as an invariant group
of this space-time. It has got two invariants: the mass and the spin,
which give the classification of all physical particles and fields.
Their field theoretical image are
the energy-momentum density tensor
and the spin density tensor.
In the general relativity the spaces $E^{(1,3)}$ are
a tangent spaces of the manifold $M^{(1,3)}$,
and one uses {\em only} the energy-momentum tensor to
define a riemannian geometry of the space-time.
It is quite natural to suppose that
the second fundamental quantity - the spin density tensor
must be used to define the whole affine-metric connection,
i.e. to define the {\em torsion}
in addition to the riemannian part of the connection.

A modern realization of this idea are the ECSK gauge theories of gravity.
They have many beautiful features, but
their present versions suffer of several difficulties.
In our opinion a very important one of them is that ECSK theories
violate the Einstein equivalence principle (EP).
According to this principle there must exist
{\em a local frame} in which the connection coefficients equals zero locally
and the motion of {\em all test particles and fields simultaneously}
looks like a free motion up to a higher order terms.

Indeed:

1) For classical  relativistic spinless particle with mass $m$
the standard application of the action principle to the action
\begin{equation}
{\cal A}= -mc^2 \int_{t_1}^{t_2} dt
\sqrt{ g_{\alpha\beta}{\dot q}^\alpha {\dot q}^\beta}
\label{eq:A1}
\end{equation}
leads to geodesic trajectories in $M^{(1,3)}$, and to dynamical equations:
\begin{equation}
{\frac {d^2 q^\gamma} {ds^2}} + { \gamma \brace \alpha \beta}
{\frac {d q^\alpha} {ds}}{\frac {d q^\beta} {ds}} = 0,
\label{eq:G}
\end{equation}
written in the proper time parametrization:
$ds = \sqrt{g_{\alpha\beta}{\dot q}^\alpha {\dot q}^\beta}dt$.

2)For  scalar field $\phi(q)$ the action
\begin{equation}
{\cal A}= {\frac 1 2} \int d^4q \left(
g^{\alpha\beta}\partial_\alpha \phi \partial_\beta \phi - m^2 \phi^2
\right)
\label{eq:APhi1}
\end{equation}
leads under standard variational principle to the equation of motion
\begin{equation}
g^{\alpha\beta}
\stackrel{\{\}}{\nabla}_\alpha \stackrel{\{\}}{\nabla}_\beta \phi
+ m^2 \phi = 0.
\label{eq:PhiG}
\end{equation}
Here $\stackrel{\{\}}{\nabla}_\alpha$ are covariant derivatives with
respect to the Levi Cevita connection with Christofel symbols
${\gamma \brace \alpha\beta}$ , which determine the riemannian
part of the affine-metric connection (\ref{eq:Gamma}),
and $g^{\alpha\beta}
\stackrel{\{\}}{\nabla}_\alpha \stackrel{\{\}}{\nabla}_\beta =
{\frac 1 {\sqrt{-g}}}\partial_\alpha(\sqrt{-g}g^{\alpha\beta}\partial_\beta)$
is the Beltrami-Laplas operator on $M(1,3)$.

Both equations (\ref{eq:G}) and (\ref{eq:PhiG}) and the corresponding
dynamics of spinless particle and spinless field do not depend on the
torsion, i.e. on the whole affine-metric connection. In contrast, the
Dirac equation for test one-half-spin-particle reads:
\begin{equation}
i\hat D \psi+ m \psi = 0.
\label{eq:PsiA}
\end{equation}

To define the Dirac operator on the affine-metric space $M(1,3)$ :
\begin{equation}
\hat D = {e^\alpha}_a \gamma^a(\partial_\alpha -
{\frac 1 4}G_{\alpha a b}\gamma^a\gamma^b)
\label{eq:D}
\end{equation}
one needs the whole connection coefficients (the "spin-connection" coefficients)
in the local anholonomic tetrad bases :
\begin{equation}
 G_{\alpha a b}  =  C_{\alpha a b} + K_{\alpha a b}.
\label{eq:Ge}
\end{equation}
Here $C_{ijk}= \Omega_{ijk}+\Omega_{kij}+\Omega_{kji}$ describe the
riemannian part of the affine-metric connection in the anholonomic bases,
${\Omega_{ij}}^k= {e^\alpha}_i {e^\beta}_j \partial_{[\alpha}{e^k}_{\beta]}$
being the components of the anholonomy object, and
$K_{\alpha a b}$ are the corresponding components of the contorsion.

We see that the one-half-spin-particle equation (\ref{eq:PsiA})
depends on the whole affine-metric connection (\ref{eq:Gamma}).
It's impossible to write down the Dirac equation on a nonflat space
in holonomic coordinate bases. Therefore,
to make the contrast between this
equation and the spin-zero-particle equations
(\ref{eq:G}) and (\ref{eq:PhiG}) transparent,
it is useful to write down the last ones in
the local tetrad bases, too.
In this bases they acquire the form:
${\frac {d^2 q^c} {ds^2}} + {C_{ab}}^c
{\frac {d q^a} {ds}}{\frac {d q^b} {ds}} = 0 $,
and
$\eta^{ab}\stackrel{\{\}}{\nabla}_a \stackrel{\{\}}{\nabla}_b \phi
+ m^2 \phi = 0$,
and these equations obviously depend only on the riemannian part ${C_{ab}}^c$
of the spin-connection, not on the contorsion ${K_{ab}}^c$,
in contrast to the Dirac equation (\ref{eq:PsiA}).

Now we see that in the ECSK theories of this type it's impossible
to find a local reference system in which both the spinless and
the spin-one-half {\em test} particles will move as a free particles.
To do this one needs to find a reference system in which both
the affine-metric connection and the Christofel symbols are zero.
This is impossible, because in presence of a nonzero torsion the difference
$ { \gamma \brace \alpha \beta} - { \Gamma_{\alpha\beta} }^\gamma=
{K_{\alpha\beta} }^\gamma$ is a nonzero {\em tensor} quantity.
Hence, it can't be made zero even at one point (q)
by any choice of the reference system.

On the other side, an equations for spinless particles and fields,
constructed by the whole affine-metric connection are
well known. These are:

1) the equation of the autoparallel lines in affine space:
\begin{equation}
{\frac {d^2 q^\gamma} {ds^2}} +  {\Gamma_{\alpha \beta}}^\gamma
{\frac {d q^\alpha} {ds}}{\frac {d q^\beta} {ds}} = 0;
\label{eq:A}
\end{equation}

2)the scalar field equation written down by the Laplas operator
$g^{\alpha\beta}\nabla_\alpha \nabla_\beta $
instead of by the Beltrami-Laplas one:
\begin{equation}
g^{\alpha\beta}
\nabla_\alpha \nabla_\beta \phi
+ m^2 \phi = 0.
\label{eq:PhiA}
\end{equation}

We shall call these equations together with the Dirac equation (\ref{eq:PsiA})
an equations of {\em autoparallel type}, and the equations
(\ref{eq:G}) and (\ref{eq:PhiG}) -- an equations of {\em geodesic type}.
All equations of autoparallel type separately obey a coherent EP
by construction. All equations of geodesic type separately do the same, too.
The problem appears when one uses an equations of both types
simultaneously, as one does in the standard EC, or ECSK theories.
Hence, to comply with EP we have to work only with one type of equations.
If we accept a geodesic type equations for  all particles,
we will lose the very idea of these theories.
Therefore, the only way to overcome the difficulty with EP in ECSK
type theories is to accept the equations of autoparallel type like
equations (\ref{eq:PsiA}), (\ref{eq:A}), and (\ref{eq:PhiA})
for all other material particles and fields, too.
The problem, one has to solve going this way is, that the {\em usual}
action principle leads just to the equations of geodesic type
for a spinless particles and fields. Hence, one needs a proper
modification of the very action principle.
In the present article we give the solution of this problem
for {\em affine flat spaces  with torsion}, as an instructive step
toward the solution for general affine-metric spaces.

\section{The Relativistic Classical Particle in \\
Affine Flat Space with Torsion}

An affine flat spaces with torsion can be produced from a flat
spaces by anholonomic transformation \cite{Kleinert1}, \cite{Kleinert2}.
We use this possibility to derive the proper form of the variational
principle in spaces with torsion performing a transformation of the standard
action principle in a flat space to the modified action principle
in a space with torsion.

It is well known that the action principle of classical mechanics  is not
invariant under anholonomic transformations. When transforming the equations
of motion anholonomicaly to new coordinates, the result does not agree with
the naively derived equations of motion of the anholonomicaly transformed
action (see \cite{FizKl1}, \cite{FizKl3}, and the references there in).

Consider the free motion of a
relativistic particle with mass $m$ in a flat Minkowski space
$E^{(1,3)} \{ {\bf x} \} \owns {\bf x} = \{ x^i\}_{i=0,1,2,3}$. The action
\begin{equation}
{\cal A}= -mc^2 \int_{t_1}^{t_2} dt
\sqrt{ \eta_{ij}{\dot x}^i {\dot x}^j}
\label{eq:A2}
\end{equation}
under standard action principle
\begin{equation}
\delta {\cal A} [{\bf x}(t) ] = 0
\label{eq:AP}
\end{equation}
leads to the dynamical equations (in the proper-time parametrization):
\begin{equation}
{\frac {d^2x^i} {ds^2}}=0
\label{eq:DE1}
\end{equation}
solved by a straight line with uniform velocity.

Let us perform the following {\em anholonomic\/} transformation
\begin{equation}
\label{eq:AT} {\frac {dx^i(s)} {ds}} =
e^i{}_\mu ({\bf q}(s)) {\frac {dq^\mu(s)} {ds}}
\end{equation}
from the Cartesian to some new coordinates ${\bf q} = \{q^ \mu \}_{ \mu
=0,1,2,3;}$, where $e^i_ \mu ({\bf q})$ are elements of some tetrads.
By assumption, they are defined at each
point of some space $M^{(1,3)} \{ {\bf q} \} \owns
{\bf q}$,  and satisfy the {\em anholonomy\/} condition
\begin{equation}
\label{eq:AC} \partial _{[ \nu } e^i{}_ {\mu]}({\bf q}) \neq 0.
\end{equation}

When inserted into the dynamical equations (\ref{eq:DE1}) the transformation
(\ref{eq:AT}) gives
$$
0={\frac {d^2x^i} {ds^2}} ={\frac d {ds}}\left( e^i{}_\mu \left({\bf q}(s)\right)
{\frac {dq^\mu} {ds}} \right) =
e^i{}_\mu \left({\bf q}(s)\right){\frac {d^2q^\mu} {ds^2}} +\partial _\nu e^i{}_\mu
\left({\bf q}(s)\right){\frac {dq^\nu} {ds}} {\frac {dq^\mu} {ds}}.
$$
After multiplying by the inverse matrix $e^\alpha {}_i({\bf q}(s))$
these equations can be written down in
a form of autoparallel equations (\ref{eq:A})
in the affine-metric space  $M^{(1,3)}\{ {\bf q; g;}\stackrel{0} {\Gamma}~\}$.
The coefficients
$\stackrel{0} {\Gamma}_{\mu \nu }{}\!\!^\alpha ({\bf q}):=
e^\alpha {}_i({\bf q})\partial _\mu e^i{}_\nu({\bf q}) $
define a flat affine connection with zero Cartan curvature
and nonzero torsion $\stackrel{0} {S}_{\mu \nu}~{}\!\!^\lambda ({\bf q}):=
\stackrel{0} {\Gamma}_{[\mu \nu]}{}\!\!^\lambda ({\bf q})\neq 0$
(because of the anholonomic condition (\ref{eq:AC})).

Another type of dynamical equations are obtained when transforming directly
the action (\ref{eq:A2}) nonholonomicaly. Under the transformation
(\ref{eq:AT}) it goes over into the action (\ref{eq:A1}).
Then the standard variational principle in the space
$M^{(1,3)}\{{\bf q;g;}\stackrel{0} {\Gamma} \}$
\begin{equation}
\label{eq:AP'}\delta {\cal A}[{\bf q}(t)]=0
\end{equation}
produces the geodesic equations of motion (\ref{eq:G}).

It is obvious that the correct equation of motion in the affine-metric space
$M^{(1,3)}\{{\bf q;g;}\stackrel{0} {\Gamma} \}$, which correspond to the
dynamical equations in the space $E^{(1,3)} \{ {\bf x} \}$
are the autoparallel equations (\ref{eq:A}), derived from equations
(\ref{eq:DE1}) by direct transformation.
Hence, something is wrong with the variational principle (\ref{eq:AP'}).

The general resolution of this conflict by an appropriate modification
of the action principle was given in (\cite{FizKl3}). For the present case
it will be based on the mappings between the tangent spaces of the
{\em anholonomic} space $E^{(1,3)} \{ {\bf x} \}$,
and {\em holonomic} one  $M^{(1,3)} \{ {\bf q} \}$,
namely the tangent map:
$$
e_{q\rightarrow x}:{\hskip 1truecm}T_{{\bf q}}{M^{(1,3)} \{ {\bf q} \}}
{\rightarrow }T_{{\bf x}}{E^{(1,3)} \{ {\bf x} \}},
$$
defined by the relations :
\begin{equation}
\left\{ 
\begin{array}{lll}
\,\,{{\bar {}\!\!d}}\!\,\hspace{0.5pt}x^i & = & e^i{}_\mu (
{\bf q})dq^\mu  \\ \,\,
{\bar {}\!\!\partial }\!\,\hspace{0.5pt}_i & = & e^\mu {}_i(%
{\bf q})\partial _\mu \,,
\end{array}
\right. 
\label{eq:x-q}
\end{equation}
and the inverse map 
$$
e_{q\leftarrow x}:{\hskip 1truecm}T_{{\bf x}}{E^{(1,3)}\{{\bf x}\}}\}
{\rightarrow }T_{{\bf q}}{M^{(1,3)} \{ {\bf q} \}}
$$
defined by the inverse relations:
\begin{equation}
\left\{ 
\begin{array}{lll}
dq^\mu  & = & e^\mu {}_i(
{\bf q})\,\,{\bar {}\!\!d}\!\,
\hspace{0.5pt}x^i \\ \partial _\mu  & = & e^i{}_\mu (%
{\bf q})\,\,\,{\bar {}\!\!\partial }\!\,\hspace{0.5pt}_i\,.
\end{array}
\right. 
\label{eq:q-x}
\end{equation}
It is easy to derive the basic relations
 $d(\,\,{\bar {}\!\!d}\!\,%
\hspace{0.5pt} x^i) = S_{jk}{}^i \,\,{\bar {}\!\!d}\!\,\hspace{0.5pt} x^j
\wedge \,\,{\bar {}\!\!d}\!\,\hspace{0.5pt} x^k,  [\, \,\,{\bar
{}\!\!\!\partial}\!\,\hspace{0.5pt}_i, \,\,{\bar {}\!\!\!\partial}\!\,%
\hspace{0.5pt}_j] = -2S_{ij}{}^k \, \,\,{\bar {}\!\!\!\partial}\!\,%
\hspace{0.5pt}_k$.

In the anholonomic case, the two mappings $e _{q \rightarrow x}$  and $e _{q
\leftarrow x}$ carry {\em only \/} the tangent spaces  of the manifolds
$M^{(1,3)}\{{\bf q} \}$  and $E^{(1,3)}\{{\bf x}\}$ into
each other. They do not specify a correspondence between the points ${\bf q}$
and ${\bf x}$ themselves. Nevertheless, the maps $e_{q \rightarrow x}$ and $%
e_{q \leftarrow x}$ can be extended  to maps of some special classes of
paths on the manifolds $M^{(1,3)}\{{\bf q}\}$  and $E^{(1,3)}\{{\bf x}\}$.
Consider the set of continuous, twice differentiable
paths $\gamma _{{\bf q}} (t): [t_1, t_2] \rightarrow M^{(1,3)}\{{\bf q}\},
\gamma _{{\bf x}} (t): [t_1, t_2] \rightarrow E^{(1,3)}\{{\bf x}\}$.
Let $C_{{\bf q}}$ and $C_{{\bf x}}$ be the corresponding
closed paths  (cycles).

The tangent maps $e _{q \rightarrow x}$ and $e _{q \leftarrow x}$ imply the
velocity maps: 
$$
\dot x^i (t) = e^i{}_ \mu ({\bf q} (t)) \dot q^ \mu (t),~~ \dot q^ \mu (t) =
e^ \mu {}_i (q(t)) \dot x ^i(t) . 
$$
If {\em in addition\/} a correspondence between {\em only two\/} points
${\bf q}_1 \in M^{(1,3)}\{{\bf q} \}$ and ${\bf x}_1\in E^{(1,3)}\{{\bf x}\}$
is specified, say ${\bf q}_1 \rightleftharpoons {\bf x%
}_1$,  then the maps $e _{q \rightarrow x}$ and $e _{q \leftarrow x}$ can be
extended to the unique maps  of the paths $\gamma _{{\bf q q}_1}(t)$ and $%
\gamma _{{\bf x x}_1} (t)$,  starting at the points ${\bf q}_1$ and ${\bf x}%
_1$, respectively. The extensions are 
\begin{equation}
\label{I} \left\{ q^ \mu (t);{\bf q} (t_1) = {\bf q}_1 \right\} {\rightarrow}
\left\{ x^i (t) = x ^i_1 + \int ^{t}_{t_1} e^i{}_ \mu ({\bf q}(t)) \dot q
^\mu (t) dt;~ {\bf x} (t_1) = {\bf x}_1 \right\} , 
\end{equation}
\begin{equation}
\label{IE} \left\{ x ^i (t);{\bf x}(t_1) = {\bf x}_t \right\} {\rightarrow}
\left\{ q^\mu (t) = q^ \mu _1 + \int_{t_1}^{t} e ^ \mu{}_i ({\bf q}(t)) \dot
x^i (t) dt;~ {\bf q} (t_1) = {\bf q}_1 \right\} . 
\end{equation}
Note the asymmetry between the two maps:  In order to find $\gamma _{{\bf x} 
{\bf x}_1} (t) = e _{q \rightarrow x} ( \gamma _{{\bf q} {\bf q}_1} (t))$ 
explicitly, one has to evaluate the integral (\ref{I}). In contrast, 
specifying $\gamma _{{\bf q} {\bf q}_1} (t) = e _{q \leftarrow x} ( \gamma _{%
{\bf x} {\bf x}_1} (t))$  requires solving the integral equation (\ref{IE}).

Another important property of the anholonomic maps $e _{q \rightarrow x}$ 
and $e _{q \leftarrow x}$ is that these do not map the cycles
(as far as the boundaries of any dimension) in one space
into the cycles (boundaries) in the other space:
$e _{q \rightarrow x}(C_{{\bf qq}_1})
\neq C_{{\bf xx}_1}$  and $e _{q \leftarrow x}(C_{{\bf x} {\bf x}_1}) \neq
C_{{\bf q} {\bf q}_1}$.
If $S_{\mu \nu}{}^{\lambda} \neq 0$
there exist, in general, a nonzero Burgers vectors:
$$
b^i[C_{{\bf q}}]: = \oint_{C_{{\bf q}}} e^i{}_\mu dq^ \mu \neq 0,
$$
and 
$$
b ^\mu [C_{{\bf x}}]: = \oint_{C_{{\bf x}}} e^{\mu}{}_i dx^i  \neq 0.
$$

Consider now the variation of the {\em paths with fixed ends in the anholonomic
space} $E^{(1,3)}\{{\bf x}\}$. Let $\gamma _{{\bf x}},
\bar \gamma _{{\bf x}} \in E^{(1,3)}\{{\bf x}\}$ are two paths
with common ends. We consider two-parametric  functions $x^i
(t, \varepsilon ) \in C^2 $ for which: $x^i (t,0) = x ^i(t), x^i (t,1) =
\bar x^i (t)$, and $x^i (t_{1,2},\varepsilon) = x^i(t_{1,2})$. Then the
infinitesimal  increment along the path is $dx^i: = \partial_t x^i (t,
\varepsilon)dt$, and the variation of the path is $\delta x^i: =
\partial_\varepsilon x^i (t, \varepsilon ) \delta \varepsilon$, with fixed
ends: $\delta x^i \vert_{t_{1,2} } =0$. We call these variations `` $\delta
_x$-variations", or "$E^{(1,3)}\{{\bf x}\}$ -space-variations".
The above definitions lead to the obvious commutation relation
$
\label{CR3} \delta _x (dx^i) - d( \delta _x x ^i ) = 0. 
$
The $e _{q \leftarrow x}$ mapping maps these paths and their variations from
the space $E^{(1,3)}\{{\bf x}\}$  to the space $M^{(1,3)}\{{\bf q}\}$.

According to our definitions, they go over into
$M^{(1,3)}\{{\bf q}\}$-paths $\gamma _{{\bf q}} = e _{q \leftarrow x} (
\gamma _{{\bf x}}), {\bar \gamma}_q  = e _{q \leftarrow x} (\bar \gamma _{%
{\bf x}})$, and we have got : the ``total $\delta _x$%
-variation" of the coordinates $q^ \mu : \, 
\delta _x q^ \mu (t):= \delta _x \left[ q^ \mu (t_1) +  \int^{t}_{t_1} e^
\mu {}_i ({\bf q}) dx^i\right]$ (this is an indirect definition accomplished
by integral  equation (\ref{IE}), the ``holonomic variation" of the
coordinates $q ^ \mu : \,\,{\bar{}\hspace{1pt}\! \!\delta }_x q^ \mu (t):=
e^ \mu{} _i ({\bf q}) \delta x^ i$, and the variations $\delta _x (dq^ \mu
): = \delta _x \left[ e^ \mu {}_i({\bf q}) dx^i\right] $. These have  the
following basic properties:
1)~$\,\,{\bar{}\hspace{1pt}\! \!\delta }_x q^ \mu (t_{1,2}) = 0$,
i.e. the fixed end condition  for {\em the holonomic\/}
$\,\,{\bar{}\hspace{1pt}\! \!\delta}_x$
--variations in
the space $M^{(1,3)}\{{\bf q} \}$ (note that {\em the total
$\delta  _x$-variations  do not possess this property\/}).  2)~$\delta _x q
^\mu (t) = \,\,{\bar{}\hspace{1pt}\! \!\delta }_x q^ \mu (t) + \Delta ^\mu 
_x (t)$, \, where
$$
\Delta ^ \mu _x (t): = \int^{t}_{t_1} d\tau \Gamma _{ \alpha \beta }{}  ^
\mu \left( dq^ \alpha \,\,{\bar{}\hspace{1pt}\! \!\delta }_x q ^\beta -
\delta _x q ^\alpha  dq^ \beta \right) 
$$
is the {\em anholonomic deviation\/}. The function $\Delta ^ \mu _x  (t)$
describes the time-evolution  of the effect of the anholonomy: $\Delta
^\mu_x (t_1) = 0$.  The final value $\Delta ^\mu_x (t_2) = b^ \mu $ is the
Burgers vector.
3)~ $ \delta _x (dq^ \mu ) - d ( \,\,{\bar{}\hspace{1pt}%
\! \!\delta }_x q^ \mu ) =  \Gamma _{ \alpha \beta }{}^ \mu \left( dq^
\alpha \,\,{\bar{}\hspace{1pt}\! \!\delta }_x q^ \beta  - \delta _x q
^\alpha dq^ \beta \right)$, or
\begin{equation}
\label{**} \delta _{x A } (dq^ \mu ) - d _A  (\,\,{\bar{}\hspace{1pt}\!
\!\delta }_x q^ \mu ) = 0. 
\end{equation}
Here $\delta _{x A} (d q^ \mu ) : = \delta _x (dq^ \mu )  + \Gamma _{ \alpha
\beta }{}^ \mu \delta_x {q^ \alpha} dq^ \beta $  is the ``absolute
variation" and $d_ A (\,\,{\bar{}\hspace{1pt}\! \!\delta }_x q^ \mu ):= d 
(\,\,{\bar{}\hspace{1pt}\! \!\delta }_x q^ \mu ) + \Gamma _{ \alpha \beta
}{}^ \mu dq^ \alpha  \,\,{\bar{}\hspace{1pt}\! \!\delta }_x q^ \beta $ is
the corresponding ``absolute differential". Combining the last two
equations we find $\delta _x(dq^ \mu) - d(\delta _x q ^ \mu) = 0 $.

The action
${\cal A}[\gamma _{{\bf x}}]= \int^{t_2}_{t_1} \Lambda({\bf\dot x},t)dt$
of a mechanical system in the space $E^{(1,3)}\{{\bf x \}}$
is mapped under $e _{q \leftarrow x}$ mapping as follows:
\begin{equation}
{\cal A}[ \gamma _{{\bf x}}] \rightarrow {\cal A} [ \gamma _{{\bf q}}] = 
\int^{t_2}_{t_1} \Lambda \left({\bf e(q)\,\dot q\,}, t\right) dt =
\int^{t_2}_{t_1} L({\bf q, \dot q,} t)dt \, .
\end{equation}

Then the variational principle
\begin{equation}
\delta _x {\cal A} [ \gamma _{{\bf q}}] =0
\label{eq:APx}
\end{equation}
implies the following  dynamical equations :
\begin{equation}
\label{DE6} \partial _ \mu L - \frac{d}{dt} (\partial _{\dot q^ \mu }L )  +
2 S_{ \nu \mu }{}^ \lambda \dot q ^ \nu  \partial _{\dot q ^ \lambda } L =0.
\label{eq:DET}
\end{equation}
The dynamical equation (\ref{eq:DET}) in presence of torsion,
generated by anholonomic transformation, may be rewritten in a form:
\begin{equation}
\frac{ \delta {\cal A} [ \gamma _{{\bf q}}]}{ \delta _x q^ \mu (t)}  = \frac{
\delta {\cal A}[ \gamma _{{\bf q}}] }{ \,\,{\bar{}\hspace{1pt}\! \!\delta }
q^ \mu (t)}  + {\cal F}_ \mu = 0. 
\end{equation}
The correct variational principle (\ref{eq:APx}) is
equivalent to a D'Alembert's type principle:
\begin{equation}
\delta _q {\cal A} [ \gamma_{{\bf q}}] + \int^{t_2}_{t_1} {\cal F}_ \mu
\delta q^ \mu = 0. 
\label{eq:Dalamber}
\end{equation}
This form involves only the usual $ M^{(1,3)} \{ {\bf q} \}$-space
variations and replaces the usual action principle (\ref{eq:AP'}).
In these equations, an additional ``torsion force"
\begin{equation}
{\cal F}_ \mu = 2 S_{ \nu \mu }{}^ \lambda \dot q  ^ \nu \partial _{\dot q^
\lambda} L
\end{equation}
appears. It is easy to show, that this force preserves the mechanical 
energy but it is a nonpotential force even in the generalized sense,
i.e. it can not be described by a new interaction term in somehow
modified lagrangian. Owing to this torsion force the equations (\ref{eq:DET})
for the case of relativistic particle in affine flat space with torsion
coincide with the autoparallel equations (\ref{eq:A}). Hence the autoparallel
type equations may be produced by action principle (\ref{eq:APx}),
or by it's equivalent form (\ref{eq:Dalamber}).

\section{The Scalar Field in Affine Flat Space with \\Torsion}

We shell derive the dynamical equation for a scalar field $\phi({\bf q})$
in an affine-metric flat space with torsion using once more an anholonomic
transformation of the action principle from the anholonomic space
$E^{(1,3)} \{ {\bf x} \}$ to the holonomic one $M^{(1,3)}\{{\bf q} \}$.
As we have seen in the previous section, the key step to the right
action principle, which produces an autoparallel-type equations of
motion for a classical particle was to impose the fixed end condition
on the variations of trajectories {\em in the anholonomic space}
$E^{(1,3)} \{ {\bf x} \}$.
Therefore in the space $M^{(1,3)} \{ {\bf q} \}$
we suppose to have a modified action principle
\begin{equation}
\delta_x {\cal A}[\phi]= 0
\label{eq:APPhiX}
\end{equation}
for scalar field $\phi$ with action ($V\subset M^{(1,3)}\{{\bf q} \}$)
\begin{equation}
{\cal A}[\phi]= \int\limits_V d^4q \sqrt{g}
{\cal L}\left(\partial \phi, \phi\right).
\label{eq:APhi2}
\end{equation}
Now in the equation (\ref{eq:APPhiX}) the variation $\delta_x$ means that
the usual boundary condition on the variation $\delta_x \phi$:
\begin{equation}
\delta_x \phi \vert_{{}_{\partial V}}= 0
\label{eq:DPhiX}
\end{equation}
must be fulfilled on some boundary in the space $E^{(1,3)} \{ {\bf x} \}$.
This leads to the condition
\begin{equation}
\int\limits_{V_x} {\bar {}\!\!d}^4x \,\,{\bar {}\!\!\partial}_i
\left( \sqrt{\eta}\,\,\partial_{ \,\,{\bar {}\!\!\partial_i \phi} }
{\cal L} \,\,\delta_x \phi \right)=
\oint_{\partial V_x} d\sigma_i
\sqrt{\eta}\,\,\partial_{ \,\,{\bar {}\!\!\partial_i \phi} }
{\cal L} \,\,\delta_x \phi = 0,
\label{eq:BoundaryConditionX}
\end{equation}
which can be written down in $M^{(1,3)}\{{\bf q} \}$-space terms
making use: 1)of the formulae (\ref{eq:x-q});
2)of the definition $\stackrel{0} {\Gamma}_{\mu \nu }{}\!\!^\alpha ({\bf q}):=
e^\alpha {}_i({\bf q})\partial _\mu e^i{}_\nu({\bf q})$
of the connection coefficients;
and 3)of it's torsion $\stackrel{0} {S}_{\mu \nu}~{}\!\!^\lambda ({\bf q}):=
\stackrel{0} {\Gamma}_{[\mu \nu]}{}\!\!^\lambda ({\bf q})\neq 0$.
Thus we reach the following form of the relation
(\ref{eq:BoundaryConditionX}):
\begin{equation}
\oint_{\partial V} d\sigma_\alpha
\sqrt{g}\,\, \partial_{\partial_\alpha \phi}
{\cal L} \,\,\delta_x \phi =
\int\limits_{V} d^4q \,\,\, \!\!\partial_\alpha
\left( \sqrt{g}\,\, \partial_{ \partial_\alpha \phi}
{\cal L} \,\,\delta_x \phi \right)=
\int\limits_{V} d^4q  \sqrt{g} \,\,2S_\alpha
 \partial_{ \partial_\alpha \phi}
{\cal L} \,\delta_x \phi.
\label{eq:BoundaryConditionQ}
\end{equation}
\\
In contrast to the case of usual $\delta\phi$ variations in the space
$M^{(1,3)}\{{\bf q} \}$ this equals zero only when
$S_\alpha := {S_{\alpha \beta}}^{\beta} = 0$.
There exists no condition of type (\ref{eq:BoundaryConditionX})
on any boundary in the space $M^{(1,3)} \{ {\bf q} \}$ if $S_\alpha \neq 0$.
Therefore, in contrast to the usual action principle,
in the space $M^{(1,3)}\{{\bf q} \}$ with a nonzero torsion vector  $S_\alpha$
the boundary terms (\ref{eq:BoundaryConditionQ})
do contribute to the variational derivative:
\begin{equation}
{ {\delta{\cal A}[\phi]} \over {\delta_x \phi} }=
{ {\delta{\cal A}[\phi]} \over {\delta \phi} } +
2S_\alpha  \partial_{ \partial_\alpha \phi} {\cal L}.
\label{eq:VD}
\end{equation}

Hence, the modified action principle (\ref{eq:APPhiX}) leads to the equation
of motion:
\begin{equation}
\partial_\phi {\cal L} -
{\frac 1 {\sqrt{g}}}
\partial_\alpha ( \sqrt{g} \partial_{\partial_\alpha \phi} {\cal L})
+ 2 S_\alpha  \partial_{ \partial_\alpha \phi} {\cal L} = 0
\label{eq:EPhi}
\end{equation}
with an additional torsion-force-density-term
$2 S_\alpha  \partial_{ \partial_\alpha \phi} {\cal L} $.

Owing to this term the modified action principle (\ref{eq:APPhiX})
is equivalent to the following D'Alembert's type principle
in the space $M^{(1,3)}\{{\bf q} \}$ with nonzero torsion:
\begin{equation}
 \delta{\cal A}[\phi] +
\int\limits_{V} d^4q  \sqrt{g} \,\,
2S_\alpha \,\, \partial_{ \partial_\alpha \phi} {\cal L}\,\, \delta\phi = 0.
\label{eq:D'AlembertPhi}
\end{equation}
For a scalar field with action (\ref{eq:APhi1}) in this space it
reproduces just the autoparallel-type equation of motion (\ref{eq:PhiA}).

\end{document}